\documentclass[
prb,twocolumn,aps]{revtex4-1}

\usepackage{graphicx}
\usepackage{dcolumn}
\usepackage{bm}
\usepackage{color}

\def\LSCO{La$_{2-x}$Sr$_x$CuO$_4$}
\def\LBCO{La$_{2-x}$Ba$_x$CuO$_4$}
\def\YBCO{YBa$_2$Cu$_3$O$_{6+y}$}

\def\C60{A$_x$C$_{60}$}

\def\BSCCO{Bi$_2$Sr$_2$CaCu$_2$O$_{8+\delta}$}

\def\HgCu3{HgCa$_2$Cu$_3$O$_{8+y}$}
\def\HgCu4{HgBa$_2$Ca$_3$Cu$_4$O$_{10+y}$}
\def\TlCu{Tl$_2$Ba$_2$CuO$_{6+\delta}$}
\def\TlCu3{Tl$_2$Ba$_2$Ca$_2$Cu$_3$O$_{10+y}$}
\def\TlCu4{Tl$_2$Ba$_2$Ca$_3$Cu$_4$O$_{12+y}$}

\def\BiCu3{Bi$_2$Sr$_2$Ca$_{2}$Cu$_3$O$_y$}

\def\8LSCO{La$_{1.88}$Sr$_{.12}$CuO$_4$}
\def\110LNSCO{La$_{1.5}$Nd$_{0.4}$Sr$_{0.1}$CuO$_{4}$}
\def\stage4LCO{La$_{2}$CuO$_{4+\delta}$}
\def\Y248{YBa$_2$Cu$_4$O$_8$}

\def\NbSe2{NbSe$_2$}
\def\TaSe2{TaSe$_2$}
\def\TiSe2{TiSe$_2$}
\def\NaCoOH2O{Na$_{0.3}$CoO$_{2y}$H$_2$O}
\def\MgB2{MgB${}_2$}

\begin{document}

\title{High Temperature Superconductivity: Ineluctable Complexity}
\author{Eduardo Fradkin}
\affiliation{Department of Physics, University of Illinois at Urbana-Champaign, Urbana, Illinois 61801-3080, USA}
\author{Steven A. Kivelson}
 \affiliation{Department of Physics, Stanford University, Stanford, CA 94305-4045, USA}
\begin{abstract}
The discovery of charge-density-wave order in the high-temperature superconductor {\YBCO} 
places charge order centre stage with superconductivity, suggesting they they are intertwined rather than competing.
\end{abstract}
\maketitle

In physics, `understanding' involves simplifying a problem as much as possible, 
but not so much that its essence is lost. 
Since  high temperature superconductivity (HTSC) was discovered  in certain copper oxide compounds in 1986$^1$, it has been widely 
accepted that the heart of the problem is the interrelation between  the ($d$-wave) superconducting state, the proximate `Mott insulating' phase that has long range antiferromagnetic order, and the mysterious 
`bad' metal phase that dominates the phase diagram at temperatures well above
Persistent evidence of  
other forms of electronic order, including charge 
and spin density wave orders  
and nematic charge order, has (mostly)  been dismissed as being an 
uninteresting  ``sideshow', a material dependent `complication,' or even a `disease'.
However, in recent years increasingly
extensive but largely indirect experimental
evidence that these other types of order
are ubiquitous in HTSC materials has
led to a slowly spreading realization
that they may have to be included in the
Ôirreducible minimumÕ to understand
HTSC. The direct observation Ñ by X-ray
diffraction Ñ of an incipient charge density
wave (CDW) in high-quality (ortho?VIII)
crystals of {\YBCO} (YBCO) has
now been independently reported by
Johan Chang et al.$^2$ in {\em Nature Physics} and
Giacomo Ghiringhelli et al.$^3$ in {\em Science}. This
is an important discovery, with the broad
implication that other forms of charge order
are inextricably intertwined with HTSC.

YBCO is the most studied HTSC
compound. It is a quasi-two-dimensional
material, as all the superconducting copper
oxides are, in which the mobile electrons
are largely confined to move in layers
made of copper and oxygen, forming a
liquid of strongly correlated degrees of
freedom carrying charge and spin quantum
numbers. A schematic phase diagram of
YBCO as a function of the temperature $T$
and the doping level $y$ (that parameterizes
the oxygen content of its chemical
composition) is shown in Fig. 1. Near
$y = 0$, and below the (N\'eel) temperature
$T_{\rm N} \sim 400$ K, YBCO is an antiferromagnet
and an electrical (Mott) insulator. As y
increases, a superconducting phase with
the shape of a rather distorted dome (red
curve) is found, which has its maximum at
the value of $T_c \sim 90$ K at the optimal doping
level of $y = 0.93$. For doping levels $y < 0.93$,
the material is considered ÔunderdopedÕ.
At temperatures well above the maximum
Tc is a strange or Ôbad metalÕ regime with
many anomalous properties that are
strikingly different from those of familiar
ÔgoodÕ metals. One of the most mysterious
regions of the phase diagram is referred
to as the Ôpseudogap regimeÕ, which lies
below a not very sharply defined crossover
temperature $T^*$, marking the boundary
between the Ôbad metalÕ and an even more
anomalous regime.
 
 \begin{figure}[t]
\begin{minipage}{0.99\linewidth}
\includegraphics[width=\linewidth]{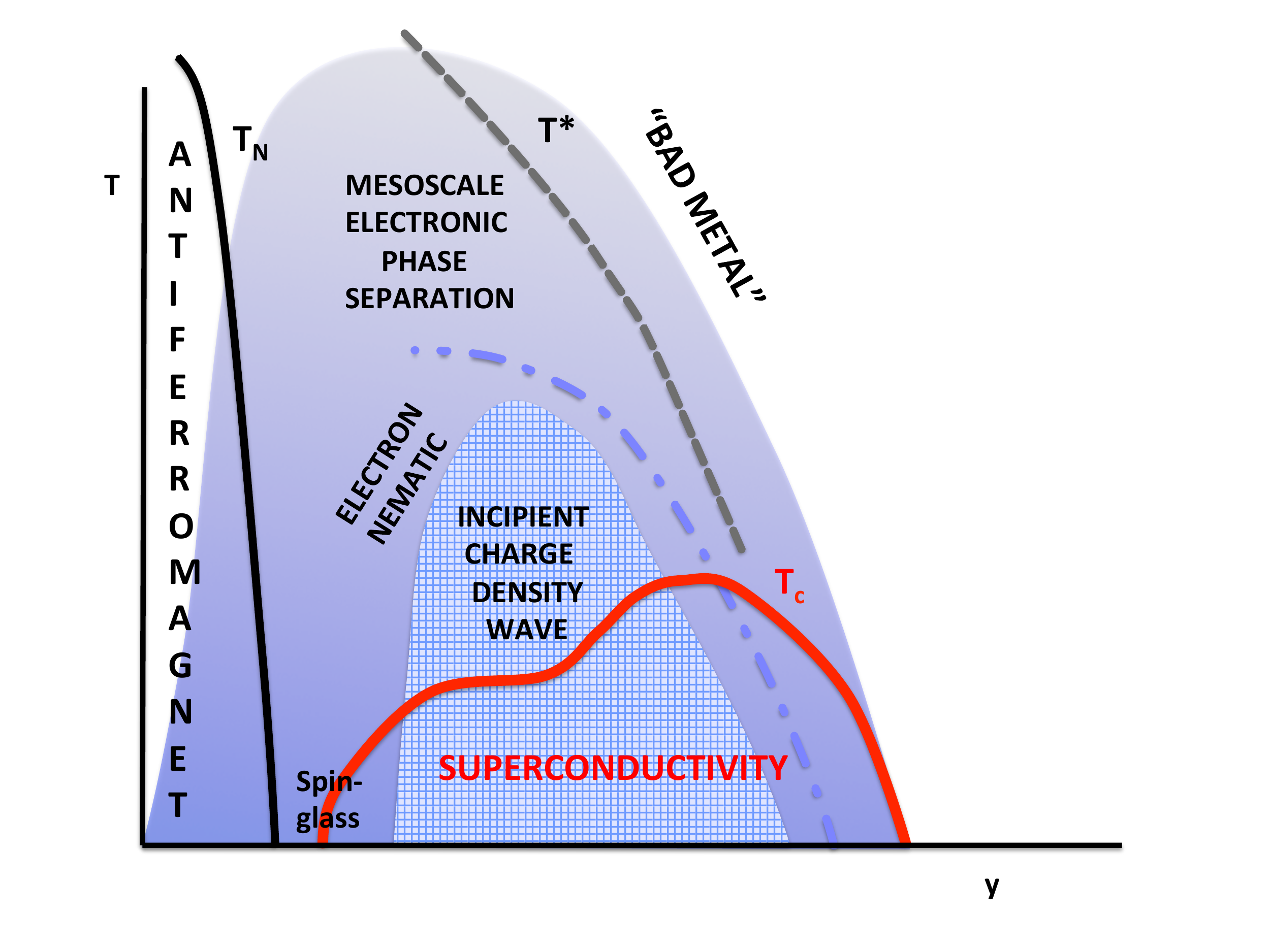}
\end{minipage}
\caption{ Schematic phase diagram of the high temperature superconductor YBCO. The red curve outlines the superconducting dome. Underneath $T^*$ lies the pseudogap regime, in which mesoscale electronic phase separation occurs.}
\label{figure}
\end{figure} 

What Chang and co-workers discovered
were pronounced peaks in the X-ray
structure factor corresponding to
substantially correlated CDW fluctuations
that emerge below a temperature,
$T_{\rm CDW} \sim 140$ K, which is lower than, but
of order $T^* \sim 250$ K, at the same level of
doping. YBCO has a crystal structure
consisting of stacked CuÐO bilayers, and
due to the presence of CuÐO chain layers.
The new peaks in the X-ray structure
factor are centred at the scattering vectors
$\textbf{Q}_1 = (q_1, 0, 1/2)$ and $\textbf{Q}_2 = (0, q2, 1/2 )$ with
$q_1 \sim q_2 \sim 0.31$. The correlation length of the
CDW order Ñ obtained from the width of
the superlattice peaks Ñ is temperature dependent,
but grows to a maximum value
corresponding to a correlation length of
roughly $\xi_{ab} \sim 20$ lattice constants in plane,
and $\xi_c ~\sim1$ unit cell out of plane. The value
of q1 implies that, in plane, the density wave
order is incommensurate with the CuÐO
lattice, with a period close to three unit cells
(which would correspond to $q_1 = 1/3$ ). The
1/2 factor in the ordering vector implies a
tendency for neighbouring bilayers to have
the CDW order shifted by phase $\pi$ relative
to each other Ñ for true long-range order,
it would correspond to a new unit cell with
four CuÐO planes.

 This fluctuating CDW order is
strongly coupled to, and competes with,
superconductivity, as demonstrated by
the observed non-monotonic temperature
dependence of the scattering intensity
and the correlation length; both grow
with decreasing temperature for
$T_{\rm CDW} > T > T_c$, then drop with further
decrease of temperature on entering the
superconducting state for $T < T_c \sim 60$ K.
This competition is further corroborated
by the high magnetic-field studies of
Chang et al.$^2$, which show that application
of magnetic fields up to 17 T perpendicular
to the planes has no detectable effect on the
CDW correlations for $T > T_c$, but produces a
large magnetic-field-induced enhancement
of both the intensity and the correlation
 length of the peak in the structure factor
when $T < T_c$. Presumably, this enhanced
CDW order is an indirect consequence
of the field-induced suppression of the
superconducting order.

Fluctuating$^3$ and static$^{5,6}$ charge- and
spin-density-wave (SDW) orders (in
the form of stripes) have been seen in
{\LSCO} and {\LBCO}, but were
regarded as special to the lanthanum
Ô214Õ family. Moreover, charge nematic
order Ñ a melted CDW/SDW state with
broken rotational invariance Ñ has been
seen in neutron scattering$^7$ in highly
underdoped YBCO (with $y = 0.45$) and also
in Nernst-effect experiments$^8$ above the
superconducting dome throughout much of
the pseudogap regime. The case of
{\LBCO}  is particularly significant as
it has a pronounced suppression of Tc near
$x = 1/8$ where static CDW and SDW (stripe)
phases are seen, as well as a state above $T_c$ in
which the CuÐO layers are superconducting
and effectively decoupled$^{9,10}$. The effective
doping of the YBCO ortho-VIII crystals
studied by Chang et al.$^2$ (also studied in
ref. 3) is also 1/8, and it is natural to expect
a relation, although the ordering wave
vectors are different. Strong evidence of
charge order (nematic and CDW) has
been seen in STM experiments$^{4,11,12}$ in
{\BSCCO}.

More directly relevant to the findings
of Chang et al.$^2$ (and to those of refs 3,4) is
the comparison with the other evidence for
charge order in YBCO. Exquisitely detailed
quantum-oscillation experiments in
magnetic fields larger than 30 T (sufficient
to suppress superconductivity in YBCO)
provided evidence that the competing
state is a density wave$^{13}$. Subsequent NMR
experiments in magnetic fields in the
range 15Ð35 T have shown that a sharp
thermodynamic phase transition occurs at
the onset of the CDW phase$^{14}$ (although the
ordering wave vector does not seem to agree
with the X-ray results). Recent ultrasound
experiments$^{15}$ have found a thermodynamic
transition at $T^*$ and a peak in the
attenuation rate at $T_{\rm CDW}$. Finally, possible
evidence of time-reversal symmetry-breaking
has been reported in YBCO with
the Kerr effect$^{16}$ and in neutron scattering$^{17}$.

Clear evidence of incipient CDW order
in YBCO is an important advance in
the field. As the ÔcleanestÕ of the cuprate
materials, any ordering tendencies
observed in YBCO are probably intrinsic.
Nonetheless, the results raise many
questions. Is the incipient CDW order
in YBCO a material-specific property of
this family of cuprates in a narrow range
of doping, or is it more ubiquitous? Is it
a close relative of the fully formed CDW
order seen in the same materials (by NMR)
at higher fields and low temperatures? Is it
closely related to the well-developed stripe
order seen in the lanthanum Ô214Õ family
and/or the short-range CDW correlations
seen on the surface of BSCCO? On a
more basic level, is the incipient CDW
order always correlated with pseudogap
formation? Is local CDW ordering the
cause of the pseudogap or a derivative
phenomenon that can, at times, arise once
the essential pseudogap correlations have
developed? Is there more to the interplay
between CDW order and HTSC than
just competition?
 
 Although there is no direct X-ray
evidence of any unidirectional (stripe)
character of the incipient CDW order in
YBCO, the fact that it occurs in a similar
location in the phase diagram and has
such similar energy and temperature
scales, suggests a single, unifying physical
significance of the CDW tendencies in
YBCO and other cuprates. In particular,
the comparable magnitudes of the
temperatures of the observed phases
 (other than antiferromagnetism) suggest
that all these orders arise together from
one ÔparentÕ state and that the various
order parameters are ÔintertwinedÕ rather
than simply competing with each other.
HTSC in the cuprates occurs on doping an
insulating antiferromagnet. In the process
of adding delocalized charge carriers
to such a strongly correlated insulating
state there is an inherent tendency of
the charge degrees of freedom (holes in
this case) to undergo electronic phase
separation$^{18}$. The combined effects of
the long-range repulsive (Coulomb)
interactions and the electronic zero-point
(quantum) kinetic energy compete with
this tendency, leading to complex phase
diagrams involving several generally
inhomogeneous and anisotropic states$^{19}$.
From this perspective, a dynamically
fluctuating, mesoscopically inhomogeneous
mixture of antiferromagnetic and
conducting (delocalized) regions may
play the role of the ÔparentÕ state out
of which the many phases, including
HTSC, emerge.
 This work was supported in part by the DOE under contracts  
DE-FG02-07ER46453 (Illinois) and  DE-AC02-76SF00515 (Stanford).

\vskip 0.5 cm

\noindent References\\
\noindent
1. Bednorz, J. G. \& M\"uller, K. A. {\em Z. Phys.B} \textbf{64}, 189-193 (1986).\\
\noindent
2. Chang J. {\it et al},  {\em Nature Phys.} \textbf{8}, 871-876 (2012).\\
\noindent
3. Ghiringhelli, G.  {\it et al}, {\it Science} {\bf 337}, 821-825 (2012).\\
\noindent
4.  Kivelson, S. A.  {\it et al},  {\it Rev. Mod. Phys.} {\bf 75}, 1201-1241 (2003).\\
\noindent
5. Tranquada, J. M.  in {\em Handbook of High-Temperature Superconductivity}(eds  Schrieffer, J. R. \&  Brooks, J.) Ch.6 (Springer, 2007).\\
\noindent
6. H\"ucker,  M.  {\it et al},  {\it Phys. Rev. B}  {\bf 83}, 104506 (2011).\\
\noindent
7. Hinkov, V.  {\it et al},  {\it Science}, {\ bf 319}, 597-600 (2008).\\
\noindent
8. Daou, R. {\it et al},  {\it Nature} \textbf{463}, 519-522 (2010).\\
\noindent
9. Li, Q.  {\it et al}, {\it Phys. Rev. Lett.} {\bf 99}, 067001 (2007).\\
\noindent
10. Berg, E. {\it et al},  {\it New J. Phys.} {\bf 11}, 115004 (2009).\\
\noindent
11. Lawler, M. J.  {\it et al},  {\it Nature} {\bf 466}, 347-351 (2010).\\
\noindent
12. Parker, C. V.  {\it et al}, {\it Nature} {\bf 468}, 677-680 (2010).\\
\noindent
13. Doiron-Leyraud, N.  {\it et al},  {\it Nature} {\bf 447},  565-568 (2007).\\
\noindent
14. Wu, T.  {\it et al}, {\it Nature} {\bf 477}, 191-194 (2011).\\
\noindent
15. Shekhter, A. {\it et al}, Preprint available at \\
 http://arxiv.org/abs/1208.5810 (2012).\\
 noindent
16. Xia, J.  {\it et al},  {\it Phys. Rev. Lett.} {\bf 100}, 127002 (2008).\\
\noindent
17. Fauqu\'e, B.  {\it et al},  {\it Phys. Rev. Lett.} {\bf 96}, 197001 (2006).\\
\noindent
18.  Emery, V. J.  \&  Kivelson, S. A. {\it Physica C} {\bf 209}, 597-621 (1993). \\
\noindent
19. Kivelson, S. A.,  Fradkin, E.  \&  Emery,, V. J. {\it Nature} {\bf 393}, 550-553 (1998).

\end{document}